\documentclass[sigconf]{acmart}

\copyrightyear{2026}
\acmYear{2026}
\setcopyright{cc}
\setcctype{by}
\acmConference[MSR '26]{23rd International Conference on Mining Software Repositories}{April 13--14, 2026}{Rio de Janeiro, Brazil}
\acmBooktitle{23rd International Conference on Mining Software Repositories (MSR '26), April 13--14, 2026, Rio de Janeiro, Brazil}
\acmPrice{}
\acmDOI{10.1145/3793302.3793616}
\acmISBN{979-8-4007-2474-9/2026/04}

\usepackage{float}
\usepackage{subcaption}
\usepackage{tabularx}
\usepackage{todonotes}
\usepackage{tcolorbox}
\usepackage{amsmath}

\usepackage{amssymb}
\usepackage{balance}
\usepackage{booktabs}
\usepackage{caption}
\usepackage{colortbl}
\usepackage{xcolor}
\usepackage{enumitem}
\usepackage{comment}
\usepackage{soul}
\usepackage{pifont}

\newcommand{\xmark}{\ding{55}}

\begin{document}

\title{Who Said CVE? How Vulnerability Identifiers Are Mentioned by Humans, Bots, and Agents in Pull Requests}

\author{Pien Rooijendijk}
\affiliation{%
  \institution{Radboud University}
  \country{The Netherlands}
}

\author{Christoph Treude}
\affiliation{%
  \institution{Singapore Management University} 
  \country{Singapore}
}

\author{Mairieli Wessel}
\affiliation{%
  \institution{Radboud University}
  \country{The Netherlands}
}

\renewcommand{\shortauthors}{Rooijendijk et al.}

\begin{abstract}
Vulnerability identifiers such as CVE, CWE, and GHSA are standardised references to known software security issues, yet their use in practice is not well understood. This paper compares vulnerability ID use in GitHub pull requests authored by autonomous agents, bots, and human developers. Using the AIDev pop dataset and an augmented set of pull requests from the same repositories, we analyse who mentions vulnerability identifiers and where they appear. Bots account for around 69.1\% of all mentions, usually adding few identifiers in pull request descriptions, while human and agent mentions are rarer but span more locations. Qualitative analysis shows that bots mainly reference identifiers in automated dependency updates and audits, whereas humans and agents use them to support fixes, maintenance, and discussion.
\end{abstract}


\maketitle

\section{Introduction}

Software vulnerabilities are routinely identified, catalogued, and communicated through standardised identifier systems such as Common Vulnerabilities and Exposures (CVE), Common Weakness Enumeration (CWE), GitHub Security Advisories (GHSA), and language specific databases such as the Go vulnerability database. These identifiers (IDs) support coordination across projects, tools, and communities by providing shared references for discussing weaknesses, tracking disclosures, and linking fixes to known issues. Prior work has shown that such IDs are referenced in open source development artefacts, including commits, issues, and pull requests, although these references often concern third-party dependencies rather than vulnerabilities in the project's own code \cite{nakano2020quantitative}.

Previous studies examine how developers respond to disclosed vulnerabilities, how long fixes take, and which weaknesses are addressed. Antal et al.~\cite{antal2020exploring} analyse commit histories and show how CWE based vulnerability types differ across ecosystems and how quickly communities react after disclosure. Kumar et al.~\cite{kumar2024comprehensive} show that vulnerable dependencies often persist and that serious vulnerabilities can remain unfixed for long periods. Kancharoendee et al.~\cite{kancharoendee2025categorizing} study vulnerability reporting on GitHub and find that vulnerabilities are often disclosed publicly despite defined private channels, affecting exposure and coordination even when fixes exist.

While prior work has used vulnerability IDs to study security related activity, their role as communication artefacts remains unclear, particularly in pull requests where humans, bots, and autonomous coding agents collaborate ~\cite{zou2025call, watanabe2025use}. Examining who mentions vulnerability IDs helps assess whether autonomous agents participate in security maintenance tasks, such as identifying, discussing, and fixing known vulnerabilities, or whether this work is still largely driven by humans and dependency management bots~\cite{alfadel2021use}.

In this paper, we study how vulnerability IDs are mentioned in GitHub pull requests. We start from the AIDev dataset \cite{li2025aidev}, which captures pull requests authored by autonomous coding agents, and augment it with additional pull requests from the same repositories but are authored by human developers or GitHub bots. This augmented dataset enables analysis beyond agent authored pull requests and allows us to examine how vulnerabilities IDs are referenced, discussed, and addressed across different contributor types. We analyse explicit references to these IDs and address the following research question: \textit{\textbf{How are vulnerabilities mentioned in pull requests?}}

Our results show that bots account for the majority of vulnerability ID mentions, typically contributing a small number of IDs per pull request and concentrating them in pull request descriptions. Human and agent mentions are less frequent overall but are more concentrated within individual pull requests and appear across a wider range of artefacts, including commit messages and titles. A qualitative analysis further reveals that bots primarily use IDs in automated dependency updates and security audits, whereas humans and agents use them to support vulnerability fixes, maintenance work, and contextual reasoning during pull request discussions.

\noindent\textbf{\textit{Data Availability}.} We made our replication package publicly available on Zenodo.\footnote{https://zenodo.org/records/18351036}

\begin{table*}[ht]
    \centering
    \small
    \begin{tabular}{lllcc}
        \hline
        \textbf{Prefix} & \textbf{Reference Database} & \textbf{Regex (for detecting a vulnerability ID mention)} & \textbf{\# AIDev-pop} & \textbf{\# Aug.} \\
        \hline
        CVE & https://cve.mitre.org & CVE$-\backslash d\{4\}-\backslash d\{2,7\}$ & 78 & 4113 \\
        GHSA & https://github.com/advisories & GHSA$-[a-z0-9]\{4\}-[a-z0-9]\{4\}-[a-z0-9]\{4\}$ & 27 & 2988 \\
        CWE & https://cwe.mitre.org & CWE$-\backslash d\{2,4\}$ & 4 & 356 \\
        GO  & https://vuln.go.dev & GO$-\backslash d\{4\}-\backslash d\{2,4\}$ & 7 & 54 \\
        RUSTSEC & https://rustsec.org  & RUSTSEC$-\backslash d\{4\}-\backslash d\{4,7\}$ & \xmark & 104 \\
        OSV & https://osv.dev/list & OSV$-\backslash d\{4\}-\backslash d\{4,7\}$ & \xmark & 5 \\
        MAL & https://github.com/ossf/malicious-packages & MAL$-\backslash d\{4\}-\backslash d\{4,7\}$ & \xmark & 1 \\
        USN & https://ubuntu.com/security/cves & USN$-\backslash d\{4\}-\backslash d\{1,2\}$ & \xmark & 1 \\
        \hline
        & & \multicolumn{1}{r}{\textbf{Total mentions per dataset}} & 116 & 7621 \\
        \hline
    \end{tabular}
    \caption{Vulnerability ID schemes and regular expressions used for detection, with counts of ID mentions in the AIDev-pop and our augmented datasets}
    \label{tab:identifiers-source}
\end{table*}

\section{Study Design}

\subsection{AIDev Dataset}
To answer our research question, we use the AIDev dataset, a large scale collection of GitHub pull requests authored by autonomous coding agents \cite{li2025aidev}. AIDev was constructed by mining pull requests associated with five agents, OpenAI Codex, Devin, GitHub Copilot, Cursor, and Claude Code, identified through agent specific signals such as account names, branch prefixes, and co-author metadata. The dataset contains 932,791 pull requests across 116,211 repositories, with data collected up to August 1, 2025.

For our specific analysis, we focus on the AIDev-pop subset, which consists of 33,596 pull requests from 2,807 repositories with more than one hundred GitHub stars. This choice increases the likelihood that pull requests are subject to active review, discussion, and moderation, which is essential for observing how vulnerability IDs are mentioned in the wild.

\subsection{Searching Vulnerability IDs}

A \textit{vulnerability identifier} (vulnerability ID) is a standardised textual label used to refer to a known software vulnerability or weakness in public databases. For example, CVE-2025-55182 is a CVE ID following the pattern CVE-YYYY-NNNNN, where the year 2025 indicates the disclosure year and the trailing number uniquely identifies the vulnerability. Similar fixed patterns are used by other ID schemes to ensure consistent referencing across tools, repositories, and development artefacts. In this study, we rely on the Open Source Vulnerability (OSV) specification,\footnote{https://ossf.github.io/osv-schema} which defines a common interchange format and supports IDs from multiple vulnerability databases and ecosystems. Although OSV includes more than twenty distinct ID schemes, our analysis of the AIDev-pop dataset revealed mentions of only four schemes. Table~\ref{tab:identifiers-source} summarises these ID schemes, together with their reference databases and the regular expressions used to detect them.

We apply these regular expressions to pull request titles, descriptions, comments, review comments, and commit messages linked to the pull request. If a single artefact contains multiple IDs, we record each ID as a separate mention. If the same ID appears in multiple places in a pull request, we count each occurrence as a mention, while counting the pull request once per vulnerability ID. For each identified vulnerability reference, we extract where in the pull request it occurred, the GitHub URL for verification, and the text in which the ID appears.

During manual inspection of the detected mentions, we removed all pull requests from the \texttt{projectdiscovery/nuclei-templates} repository. This repository contains vulnerability detection templates rather than vulnerability fixes or discussion, and produced a large number of false positives. One pull request alone\footnote{https://github.com/projectdiscovery/nuclei-templates/pull/12521} updated 3,147 CVE templates, mentioning 120 unique CVE IDs in the pull request body. Including this repository would have disproportionately skewed the results. 

\subsection{Augmenting the Dataset}

AIDev-pop provides complete pull request metadata only for agent authored contributions. To include pull requests that were not created by agents, we augment the dataset by collecting additional pull requests from all repositories in the AIDev-pop subset, restricting the collection to pull requests that explicitly mention vulnerability IDs. For each repository, we query GitHub using the GitHub REST API and apply the ID detection approach described in Section 2.2 across pull request artefacts. 
At the time of data collection, some repositories were no longer available, resulting in a final set of 2,793 repositories that were searched for vulnerability ID mentions. The augmented search is restricted to the same time period as the AIDev dataset, from January 1, 2025 to August 1, 2025. We make this augmented dataset available in our replication package.

\subsection{Identifying Bots}

We identify bot activity using both the AIDev-pop and the augmented dataset by first constructing a list of all unique accounts that reference vulnerability IDs in pull requests, including pull request authors, committers, commenters, and reviewers. We then classify each account to determine whether it corresponds to a bot, a human user, or one of the five autonomous agents included in AIDev-pop, which may also appear in the augmented dataset.

Our bot identification approach is based on three high precision signals commonly used in prior work \cite{bothunter,wessel2018,chidambaram2023dataset}: GitHub accounts explicitly tagged as bots, for example \texttt{dependabot[bot]}, usernames containing the string “bot”, and a curated list of 385 well known bot accounts reported by Chidambaram et al.~\cite{chidambaram2023dataset}. We identified 48 distinct bot accounts, all of which were manually double checked to increase the reliability of our identification. The five most frequent bot accounts are \texttt{dependabot[bot]} with 3,906 occurrences, \texttt{renovate[bot]} with 899 occurrences, \texttt{coderabbitai[bot]} with 72 occurrences, \texttt{socket-security[bot]} with 69 occurrences, and \texttt{mergify[bot]} with 52 occurrences.

\subsection{Data Analysis}
We generate descriptive statistics over the AIDev-pop subset and our augmented dataset to quantify the distribution, location, and frequency of vulnerability ID mentions across contributor types.
To further understand how vulnerability IDs are used in practice, we conduct a qualitative analysis of pull requests containing ID mentions. For this analysis, we examined each pull request directly on GitHub and read its content to understand why the ID was mentioned. Two authors independently coded a subset of 70 randomly chosen pull requests to identify recurring usage patterns and contexts, discussed disagreements to agree on a shared coding scheme.

\section{Findings}

\begin{figure}
    \centering
    \includegraphics[width=8cm]{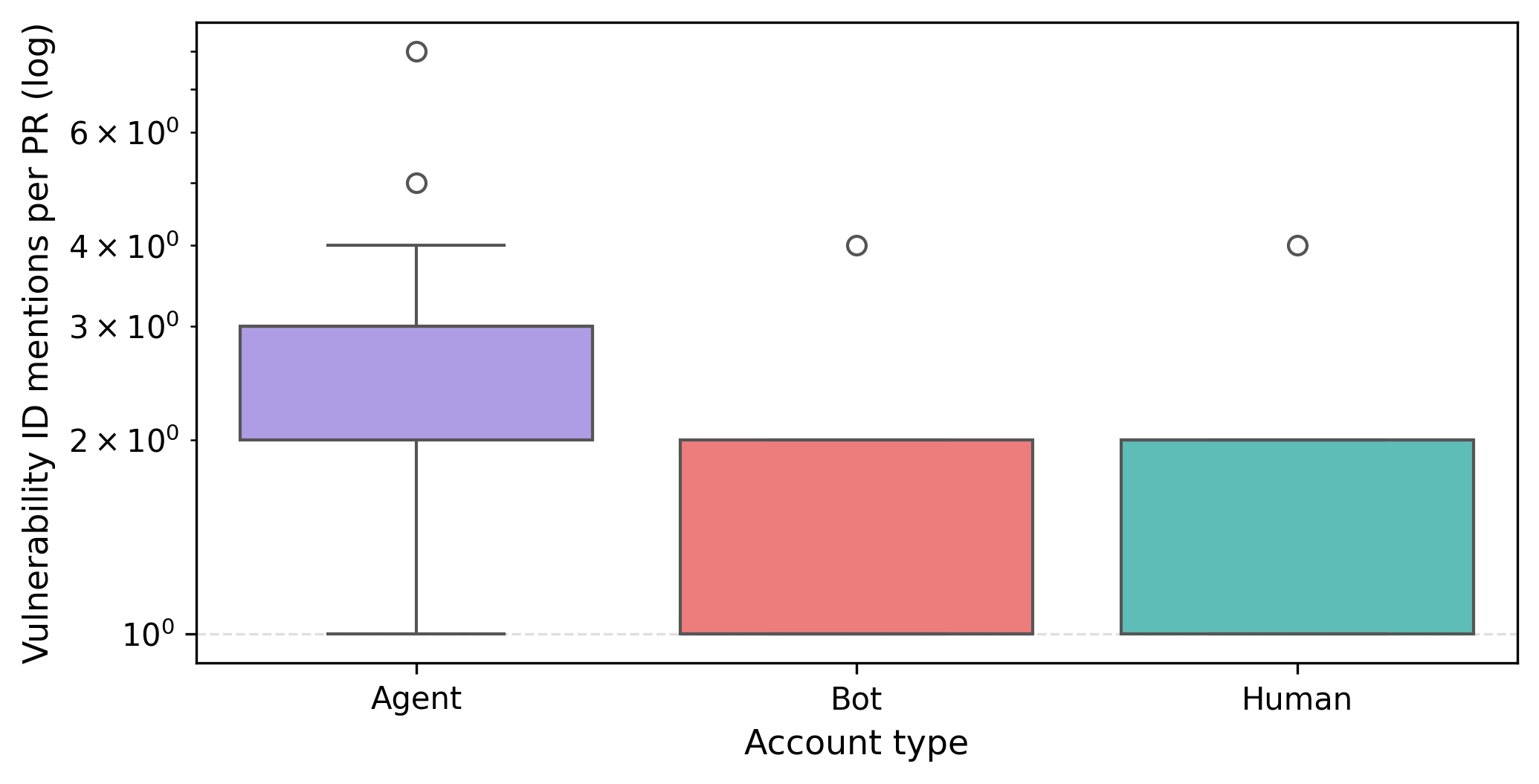}
    \caption{Distribution of vulnerability ID mentions per pull request by account type (log scale)}
    \label{fig:median_mention_per_actor_per_pr}
\end{figure}

Bot accounts account for 5,345 vulnerability ID mentions across 4,184 distinct pull requests, followed by human accounts with 2,302 mentions across 1,247 pull requests, and agent accounts with 90 mentions across 34 pull requests. Despite these differences in volume, vulnerability IDs are typically mentioned only a small number of times within an individual pull request. The median number of mentions per pull request is one for bots and humans, and two for agents. Figure~\ref{fig:median_mention_per_actor_per_pr} shows the distribution of vulnerability ID mentions per pull request by account type.

Human accounts exhibit a broader distribution of mentions within individual pull requests. While 627 pull requests contain a single vulnerability ID mention by a human account, many include multiple mentions, including 393 pull requests with two mentions and 146 with three mentions. A small number of pull requests contain comparatively high numbers of mentions by human accounts, with isolated cases reaching up to 22 and 51 mentions within a single pull request. One OWASP pull request\footnote{https://github.com/OWASP/mastg/pull/3151} accounted for the 22 mentions during a discussion on updating the mapping between another security weakness ID (MASWE) and CWE IDs to improve classification accuracy. Similarly, the pull request with 51 mentions was associated with improvements to the Github CodeQL tool,\footnote{https://github.com/github/codeql/pull/19957} where these IDs were referenced to ensure consistent detection and reporting of security issues across multiple programming languages.

\begin{figure}
    \centering
    \includegraphics[width=8cm]{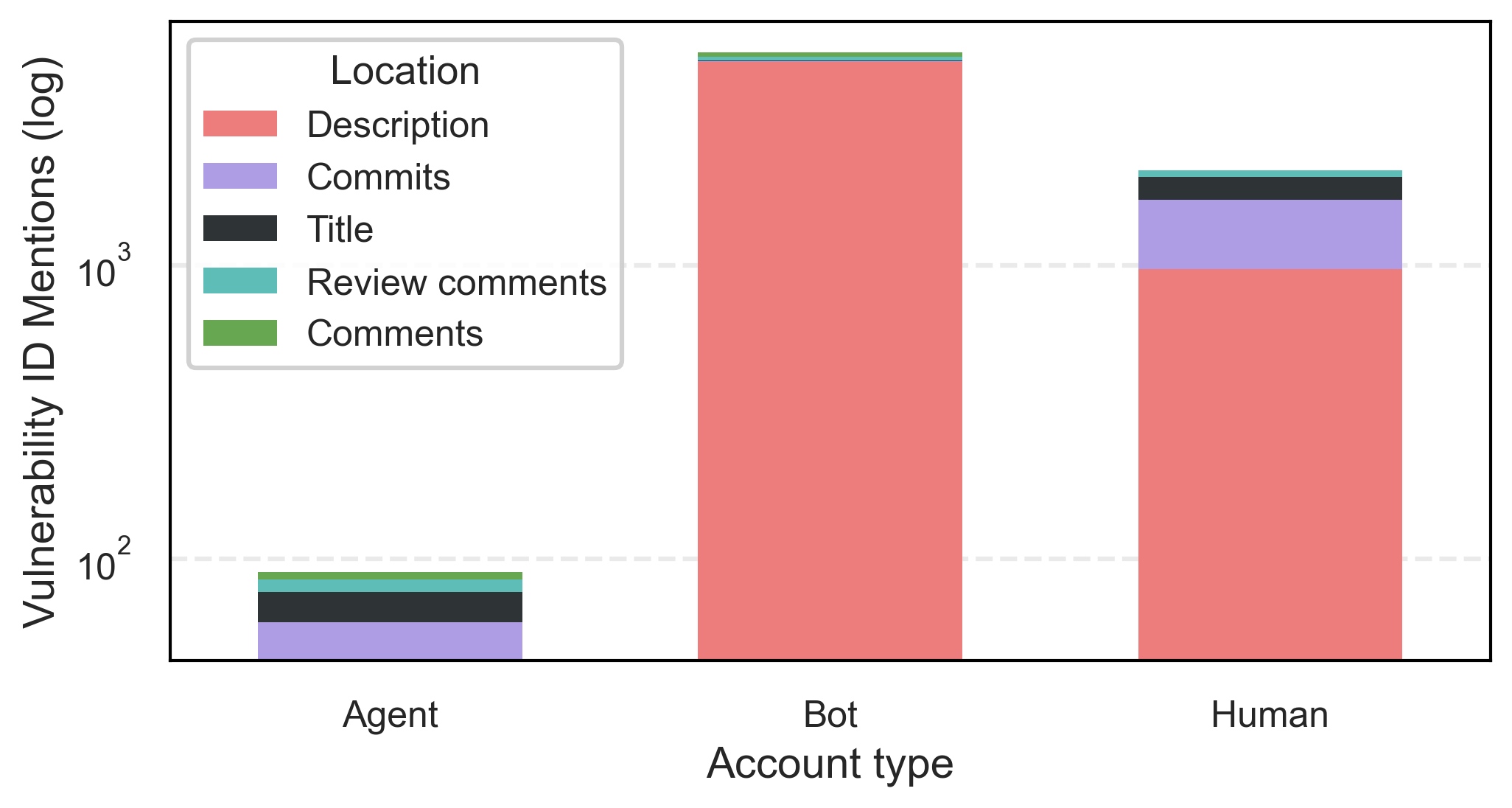}
    \caption{Distribution of vulnerability ID mentions by account type and location (log scale)}
    \label{fig:actor_type_by_mention_location}
\end{figure}

Figure~\ref{fig:actor_type_by_mention_location} shows where vulnerability IDs are mentioned within pull requests, grouped by account type. Bot accounts mention IDs almost exclusively in pull request descriptions, accounting for 4,938 mentions, while mentions in commits, titles, comments, and review comments are comparatively rare. Human accounts also most frequently mention IDs in pull request descriptions (971 mentions), but additionally refer to them in commit messages (700 mentions) and pull request titles (338 mentions). Agent accounts mention vulnerability IDs far less often overall; when they do, mentions appear mainly in pull request descriptions (45), commit messages (16), and titles (16), with very few occurrences in comments or review comments.

While these results quantify who mentions vulnerability IDs and where they appear, they do not explain the intent behind these mentions. To address this, we qualitatively analysed the contexts in which IDs are used and identified several recurring reasons for mentioning vulnerability IDs. Bots predominantly reference IDs during automated dependency updates and security audits. Humans most often mention IDs when applying or refining vulnerability fixes, resolving merge conflicts, or engaging in pull request discussions. Agent mentions reflect a mixed pattern, spanning both mitigation activities and automated security checks.

\paragraph{\textbf{Automated dependency update} (47 occurrences).}
In this category, vulnerability IDs appear in pull requests that update dependencies to versions containing security fixes. These mentions are typically generated by dependency management bots, such as \texttt{dependabot[bot]} and \texttt{renovate[bot]}, and primarily serve a referential purpose by pointing maintainers to the relevant security advisory. For example, a pull request submitted by \texttt{dependabot[bot]}\footnote{https://github.com/novuhq/novu/pull/7552} notes that a dependency update ``contains a breaking change due to security fixes'' and explicitly links to the corresponding advisory (``See GHSA-vg6x-rcgg-rjx6 for more details''). In some cases, IDs also appear in brief review exchanges, where contributors ask for confirmation that a given CVE is fixed in the proposed version, and an automated assistant responds\footnote{https://github.com/microsoft/azurelinux/pull/13893/files\#r2109724776} by pointing to the official advisory and suggesting that the pull request description include this reference to improve review clarity and future traceability.

\paragraph{\textbf{Vulnerability fix or mitigation} (12 occurrences).}
Here, vulnerability IDs are mentioned to indicate that a change directly addresses a known security issue. The ID serves as a concise signal that a pull request applies a remediation or patch, for example through commit messages or pull request titles such as ``Addresses CVE-2024-3567.''\footnote{https://github.com/microsoft/azurelinux/pull/13449} In other cases, IDs are used to justify or explain concrete code changes during review. For instance, when a contributor asks an automated assistant why a dependency was updated, the assistant explains that the change was made to fix a moderate security vulnerability (CVE-2023-36563),\footnote{https://github.com/microsoft/powerplatform-vscode/pull/1202/files\#r2108167926} describing the issue and how the update resolves it while maintaining compatibility.

\paragraph{\textbf{Dependency security audit} (10 occurrences).}
In this category, vulnerability IDs are mentioned as the outcome of automated security analysis or verification activities. These mentions are typically produced by bots and report detected security issues rather than proposing fixes directly. For example, in a pull request comment generated by \texttt{coderabbitai[bot]}, a hard-coded credential is flagged during analysis, with the comment explicitly referencing the corresponding weakness ID (CWE-798) and linking to external documentation.\footnote{https://github.com/tramlinehq/tramline/pull/855} The ID is used to classify the detected issue and support the reported security finding.

\paragraph{\textbf{Reference to advisory or illustrative example} (1 occurrence).}
Finally, vulnerability IDs are sometimes used as comparative or illustrative references rather than to report or fix a vulnerability in the code under review. Contributors draw on IDs from related systems to reason about potential risks and design trade-offs. For example, in a pull request discussion,\footnote{https://github.com/gofiber/fiber/pull/3508\#issuecomment-3105490792} a contributor references a vulnerability from another context to assess the relevance, exploitability, and potential impact of a similar design choice, even though the referenced vulnerability does not directly apply to the project itself.

\section{Threats to Validity}
Our analysis relies on explicit textual mentions of vulnerability IDs, which means we miss cases where vulnerabilities are addressed without naming an ID. Bot and agent identification is based on GitHub labels, username patterns, and a curated list of known bot accounts, which has high precision but may still misclassify some accounts. The qualitative coding of pull requests may introduce subjectivity, which we reduce by independent coding and discussion between authors. Finally, our study focuses on repositories in the AIDev-pop subset and on pull requests from January 1 to August 1, 2025, so the findings may not generalise to smaller or private repositories, other platforms, earlier periods, or ecosystems that use different vulnerability ID schemes.

\section{Related work}

Research on software vulnerabilities has long relied on standardised IDs study security practices in open source development. Early work showed that developers explicitly reference CVEs in commits, issues, and bug reports, using these IDs to signal security relevance and to coordinate fixes across repositories and ecosystems \cite{nakano2020quantitative}. Antal et al.~\cite{antal2020exploring} analysed vulnerability-resolution commits across Python and JavaScript projects, using CWE IDs to characterise vulnerability types and to measure delays between disclosure and mitigation. Other work has examined the process implications of referencing known vulnerabilities. Bühlmann and Ghafari showed that security issues explicitly referencing vulnerabilities tend to be resolved faster, suggesting that IDs support prioritisation and coordination \cite{buhlmann2022developers}. Horawalavithana et al.~\cite{horawalavithana2019cross} analysed vulnerability mentions across social media platforms and GitHub, demonstrating that CVE IDs act as stable tokens linking discussions across platforms, although without analysing how they are used within pull request.

Automation has received increasing attention, particularly in studies of dependency management bots \cite{mohayeji2025securing}. Alfadel et al.~\cite{alfadel2021use} analysed Dependabot-generated security pull requests at scale, focusing on adoption, merge rates, and time-to-merge. Rojpaisarnkit et al.~\cite{rojpaisarnkit2024} studied contributions coinciding with vulnerability mitigation in npm libraries using advisory metadata to define mitigation periods. These studies establish that bot-generated security pull requests are common and impactful, but they do not examine how vulnerability IDs are expressed in pull requests.

\section{Conclusion}
\textbf{\textit{Who said CVE?}} Vulnerability IDs are used differently depending on the actor, serving distinct coordination roles in pull requests. Bot generated mentions primarily function as lightweight signals that justify automated dependency updates or report audit findings, often confined to pull request descriptions. In contrast, human and agent mentions are more tightly integrated into the development process, appearing in commit messages, titles, and discussions where they are used to explain fixes, reason about risk, or support design decisions. As autonomous agents increasingly participate in security related development, these differences suggest a need for tooling and review workflows that better support contextual ID usage beyond automated signalling, for example by preserving rationale, linking IDs to specific code changes, and making security intent more visible to reviewers. Future work could examine how these different forms of ID usage influence review outcomes, remediation timelines, and trust in automated contributions, as well as whether similar patterns hold across other ecosystems and collaboration platforms.

\begin{acks}
Pien's work is supported by the Dutch science foundation NWO through the KIC ``Find2Fix'' project (No. KICH1.VE05.23.008).
Special thanks to Fred \includegraphics[height=0.9em]{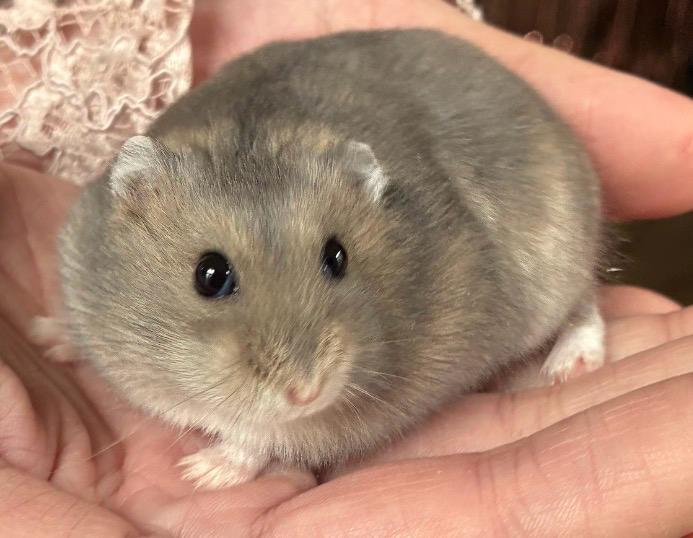} and Milo \includegraphics[height=1em]{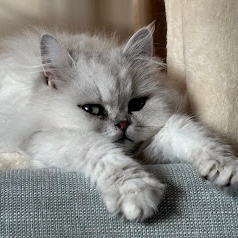}.
\end{acks}

\bibliographystyle{ACM-Reference-Format}
\bibliography{main}

@article{li2025aidev,
title={{The Rise of AI Teammates in Software Engineering (SE) 3.0: How Autonomous Coding Agents Are Reshaping Software Engineering}}, 
author={Li, Hao and Zhang, Haoxiang and Hassan, Ahmed E.},
journal={arXiv preprint arXiv:2507.15003},
year={2025}
}

@inproceedings{alfadel2021use,
  title={On the use of dependabot security pull requests},
  author={Alfadel, Mahmoud and Costa, Diego Elias and Shihab, Emad and Mkhallalati, Mouafak},
  booktitle={2021 IEEE/ACM 18th International conference on mining software repositories (MSR)},
  pages={254--265},
  year={2021},
  organization={IEEE}
}

@inproceedings{antal2020exploring,
  title={Exploring the security awareness of the python and javascript open source communities},
  author={Antal, G{\'a}bor and Keleti, M{\'a}rton and Heged{\u{u}}s, P{\'e}ter},
  booktitle={Proceedings of the 17th International Conference on Mining Software Repositories},
  pages={16--20},
  year={2020}
}

@article{nakano2020quantitative,
  title={A Quantitative Study of Security Bug Fixes of GitHub Repositories},
  author={Nakano, Daito and Yin, Mingyang and Sato, Ryosuke and Hindle, Abram and Kamei, Yasutaka and Ubayashi, Naoyasu},
  journal={arXiv preprint arXiv:2012.08053},
  year={2020}
}

@article{wessel2018,
author = {Wessel, Mairieli and de Souza, Bruno Mendes and Steinmacher, Igor and Wiese, Igor S. and Polato, Ivanilton and Chaves, Ana Paula and Gerosa, Marco A.},
title = {The Power of Bots: Characterizing and Understanding Bots in OSS Projects},
year = {2018},
issue_date = {November 2018},
publisher = {Association for Computing Machinery},
address = {New York, NY, USA},
volume = {2},
number = {CSCW},
url = {https://doi.org/10.1145/3274451},
doi = {10.1145/3274451},
journal = {Proc. ACM Hum.-Comput. Interact.},
month = nov,
articleno = {182},
numpages = {19}
}

@inproceedings{bothunter,
author = {Abdellatif, Ahmad and Wessel, Mairieli and Steinmacher, Igor and Gerosa, Marco A. and Shihab, Emad},
title = {BotHunter: an approach to detect software bots in GitHub},
year = {2022},
isbn = {9781450393034},
publisher = {Association for Computing Machinery},
address = {New York, NY, USA},
url = {https://doi.org/10.1145/3524842.3527959},
doi = {10.1145/3524842.3527959},
booktitle = {Proceedings of the 19th International Conference on Mining Software Repositories},
pages = {6–17},
numpages = {12},
location = {Pittsburgh, Pennsylvania},
series = {MSR '22}
}

@INPROCEEDINGS{rojpaisarnkit2024,
  author={Rojpaisarnkit, Ruksit and Damrongsiri, Hathaichanok and Treude, Christoph and Ouni, Ali and Kula, Raula Gaikovina},
  booktitle={2024 IEEE/ACIS 22nd International Conference on Software Engineering Research, Management and Applications (SERA)}, 
  title={Characterising Contributions that Coincide with Vulnerability Mitigation in NPM Libraries}, 
  year={2024},
  volume={},
  number={},
  pages={237-242},
  doi={10.1109/SERA61261.2024.10685587}}

@inproceedings{horawalavithana2019cross,
author = {Horawalavithana, Sameera and Bhattacharjee, Abhishek and Liu, Renhao and Choudhury, Nazim and O. Hall, Lawrence and Iamnitchi, Adriana},
title = {Mentions of Security Vulnerabilities on Reddit, Twitter and GitHub},
year = {2019},
isbn = {9781450369343},
publisher = {Association for Computing Machinery},
address = {New York, NY, USA},
url = {https://doi.org/10.1145/3350546.3352519},
doi = {10.1145/3350546.3352519},
booktitle = {IEEE/WIC/ACM International Conference on Web Intelligence},
pages = {200–207},
numpages = {8},
location = {Thessaloniki, Greece},
series = {WI '19}
}

@inproceedings{buhlmann2022developers,
  title={How do developers deal with security issue reports on github?},
  author={B{\"u}hlmann, Noah and Ghafari, Mohammad},
  booktitle={Proceedings of the 37th ACM/SIGAPP Symposium on Applied Computing},
  pages={1580--1589},
  year={2022}
}

@inproceedings{kumar2024comprehensive,
  title={A Comprehensive Study on the Impact of Vulnerable Dependencies on Open-Source Software},
  author={Kumar, Shree Hari Bittugondanahalli Indra and Sampaio, L{\'\i}lia Rodrigues and Martin, Andr{\'e} and Brito, Andrey and Fetzer, Christof},
  booktitle={2024 IEEE 35th International Symposium on Software Reliability Engineering (ISSRE)},
  pages={96--107},
  year={2024},
  organization={IEEE}
}

@inproceedings{chidambaram2023dataset,
  title={A dataset of bot and human activities in GitHub},
  author={Chidambaram, Natarajan and Decan, Alexandre and Mens, Tom},
  booktitle={2023 IEEE/ACM 20th International Conference on Mining Software Repositories (MSR)},
  pages={465--469},
  year={2023},
  organization={IEEE}
}

@article{mohayeji2025securing,
  title={Securing dependencies: A comprehensive study of Dependabot’s impact on vulnerability mitigation},
  author={Mohayeji, Hamid and Agaronian, Andrei and Constantinou, Eleni and Zannone, Nicola and Serebrenik, Alexander},
  journal={Empirical Software Engineering},
  volume={30},
  number={3},
  pages={89},
  year={2025},
  publisher={Springer}
}

@inproceedings{kancharoendee2025categorizing,
  title={On Categorizing Open Source Software Security Vulnerability Reporting Mechanisms on GitHub},
  author={Kancharoendee, Sushawapak and Phichitphanphong, Thanat and Jongyingyos, Chanikarn and Reid, Brittany and Kula, Raula Gaikovina and Choetkiertikul, Morakot and Ragkhitwetsagul, Chaiyong and Sunetnanta, Thanwadee},
  booktitle={2025 IEEE International Conference on Software Analysis, Evolution and Reengineering (SANER)},
  pages={751--756},
  year={2025},
  organization={IEEE}
}

@article{zou2025call,
  title={A Call for Collaborative Intelligence: Why Human-Agent Systems Should Precede AI Autonomy},
  author={Zou, Henry Peng and Huang, Wei-Chieh and Wu, Yaozu and Miao, Chunyu and Li, Dongyuan and Liu, Aiwei and Zhou, Yue and Chen, Yankai and Zhang, Weizhi and Li, Yangning and others},
  journal={arXiv preprint arXiv:2506.09420},
  year={2025},
  doi={https://doi.org/10.48550/arXiv.2506.09420}
}

@article{watanabe2025use,
  title={On the use of agentic coding: An empirical study of pull requests on github},
  author={Watanabe, Miku and Li, Hao and Kashiwa, Yutaro and Reid, Brittany and Iida, Hajimu and Hassan, Ahmed E},
  journal={arXiv preprint arXiv:2509.14745},
  year={2025}
}

\end{document}